\documentstyle[12pt,world_sci]{article}

\begin{document}
\begin{flushright}
FERMILAB-Conf-94/322-T\\
hep-ph/9409312\\
September 13, 1994\\
\end{flushright}
\vspace{-0.3in}
\title{{\bf SUMMARY OF TOP QUARK PHYSICS}
{\thanks {Presented by at DPF'94, Univ. of New Mexico, Albuquerque,
New Mexico, August 2-6, 1994.} }
}
\author{STEPHEN PARKE \\
{\em parke @ fnal.gov \\
Department of Theoretical Physics\\
Fermi National Accelerator Laboratory\\
Batavia, Illinois  60510, USA}\\}

\maketitle
\setlength{\baselineskip}{2.6ex}

\begin{abstract}
I briefly review standard top quark physics at
hadron colliders
and summarize the contributions to this conference.
The possibility of new mechanisms for $t\bar{t}$ production are also
discussed.

\end{abstract}


\section{Standard}
In hadron colliders the dominant mode of top quark production is via
quark-antiquark annihilation or gluon-gluon fusion,
\begin{eqnarray*}
q~\bar{q} & \rightarrow & t ~\bar{t} \\
g~g  & \rightarrow & t ~\bar{t}.
\end{eqnarray*}
However there are other
modes,
\begin{eqnarray*}
W^+ ~g  & \rightarrow & t ~\bar{b} \\
W^{+*}  & \rightarrow & t ~\bar{b} \\
(\gamma,Z) ~g  & \rightarrow & t ~\bar{t} \\
(\gamma,Z)^*  & \rightarrow & t ~\bar{t} \\
& \cdots &.
\end{eqnarray*}
In this list I have not included processes which pick
a $b$-quark out of the hadron.

These processes are approximately ordered according to
their rates in hadron colliders. Fig. 1(a) has the rates for the first three
processes at the Tevatron assuming that the dominant decay model for the top
quark is $W^+ ~b$. The channel, positron plus jets, was chosen so that
the final state for all three processes is positron, $b \bar{b}$ plus jets.
The QCD, $W$-gluon and $W^*$ processes have two, one and zero
non-$b$-quark jets, respectively.

\begin{figure}[t]
\vspace{8cm}
\includegraphics{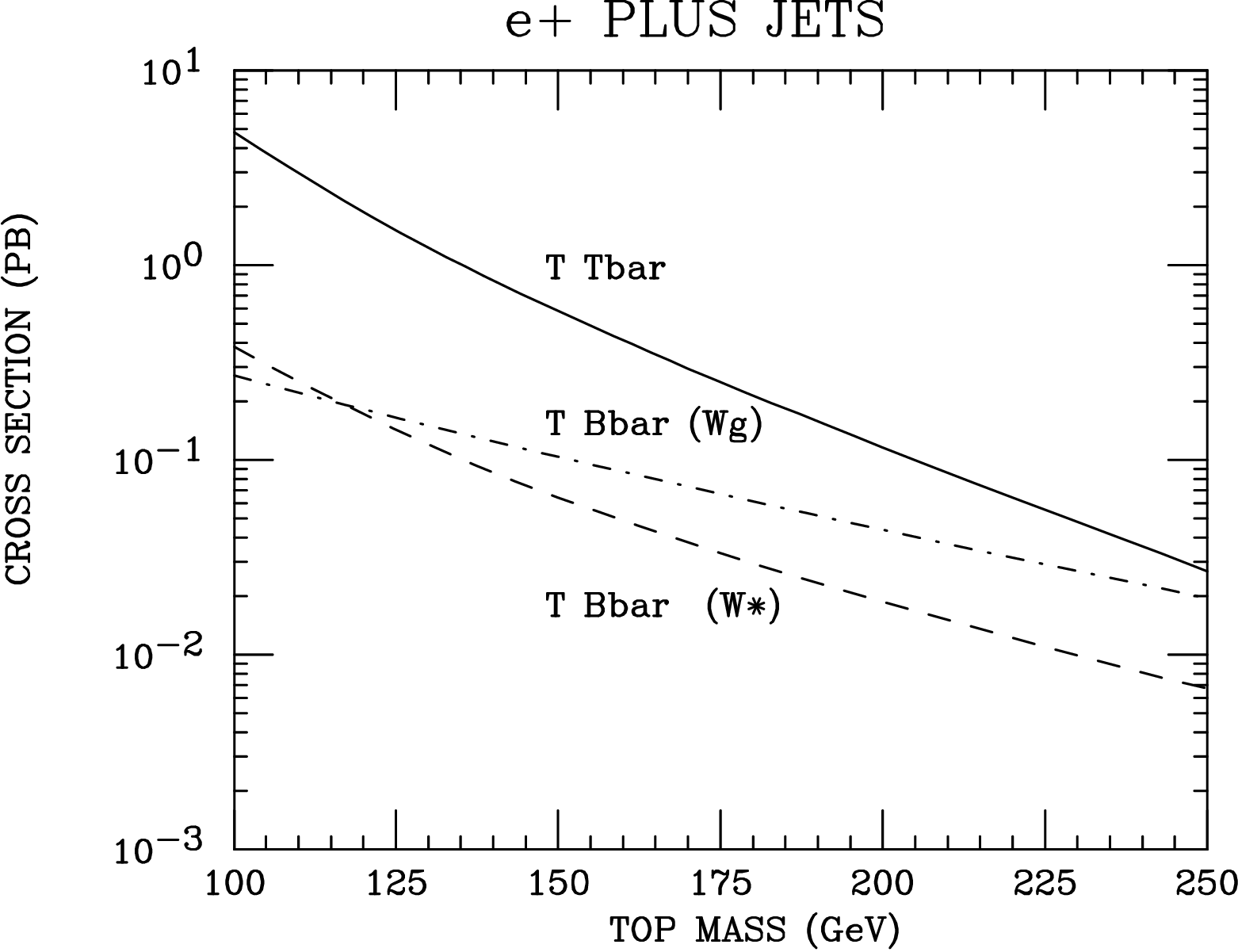}
\includegraphics{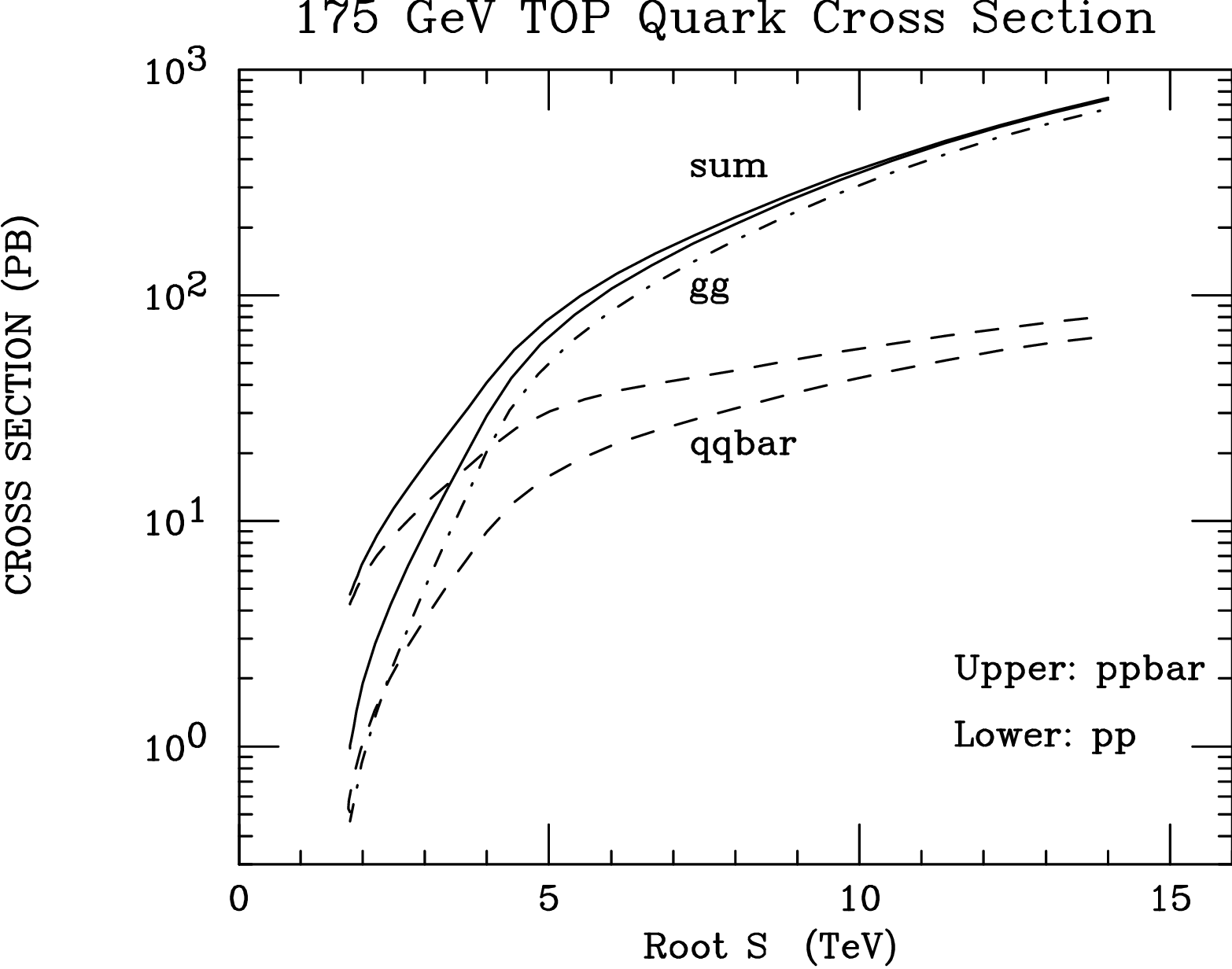}
\vspace{-1.5cm}\hspace{2cm}(a)\hspace{7cm}(b)\vspace{1cm}
\vspace{-1cm}
\caption[]{(a) The Top Quark Production Cross Section at the Tevatron.
The three curves are for quark-antiquark annihilation plus
gluon-gluon fusion (solid),
W-gluon fusion (dot-dash) and through an off-mass shell W-Boson (dashes).
(b) QCD Top quark Production cross section as a function of $\sqrt{s}$,
for quark-antiquark annihilation (dashes), gluon-gluon fusion (dot-dash)
and the sum (solid)  for both proton-antiproton (upper)
and proton-proton (lower) colliders.}
\label{xsec}
\end{figure}

\begin{figure}[t]
\vspace{8cm}
\includegraphics{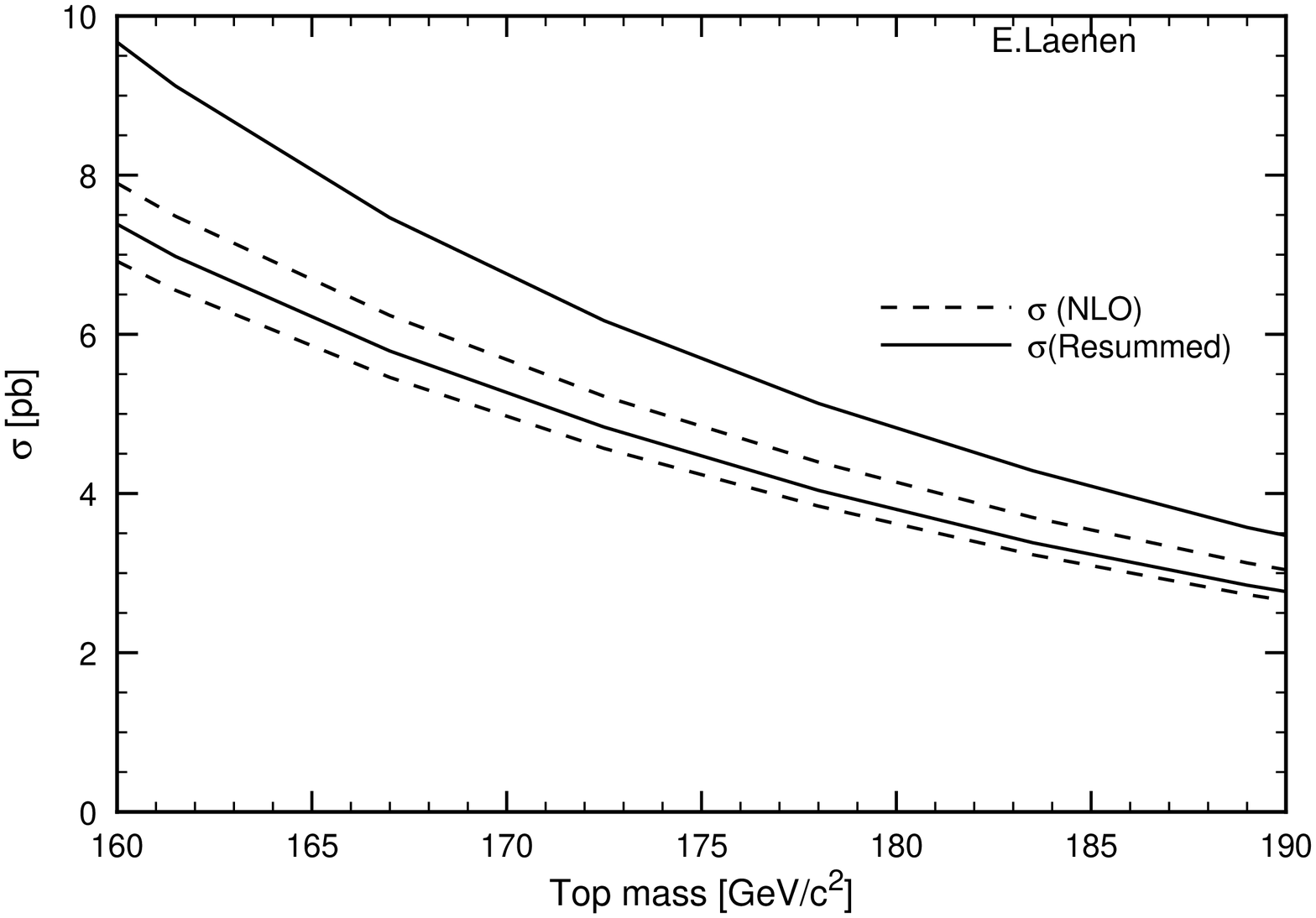}
\vspace{-1cm}
\caption[]{The QCD Top Quark Production Cross Section at the Tevatron.
The solid curves give the range of values
using the resummed next to leading order
calculations and the dashed curves are the range
for the next to leading oder calculations.
The same structure functions were used.}
\label{xsecel}
\end{figure}

Fig. 1(b) contains the QCD cross section for $m_{t} ~=~ 175~GeV$ verses
$\sqrt{s}$
for both proton-proton colliders and for proton-antiproton colliders.
At $\sqrt{s} ~=~ 1.8~TeV$ the gluon-gluon fusion is only 10\% of the cross
section for a proton-antiproton collider, the Tavatron,
whereas at a $14~TeV$ proton-proton
collider, the LHC, gluon-gluon fusion is 90\% of the cross section.

For an accurate determination of the QCD top cross section,
we need to consider the next to
leading order calculations and the soft gluon resummation of
the next to leading order calculations\cite{epl}.
In Fig. 2, the results of these calculations using the same
structure functions are shown.
At high top quark masses the
difference between these two calculations is at the 20\% level.

The standard model decays of the top are
\begin{eqnarray*}
t & \rightarrow & W^+ ~b \\
t & \rightarrow & W^+ (~s ~or~d) \\
t & \rightarrow & g ~W^+ (~b, ~s ~or~d) \\
t & \rightarrow & \gamma ~W^+ (~b, ~s ~or~d) \\
t & \rightarrow & Z ~W^+ (~b, ~s ~or~d) \\
t & \rightarrow & \phi ~W^+ (~b, ~s ~or~d) \\
& \cdots &.
\end{eqnarray*}
For $m_t ~=~175 ~GeV$, the total width of the
top quark is approximately $1.5~GeV$.
The CKM suppressed decays are expected to be less than 0.1\% of the
non-suppressed decays.
Whereas the decays including a $Z$ or Higgs
will be extremely small unless the on-mass shell decay is kinematically
allowed.

Flavor changing neutral currents,
\begin{eqnarray*}
t & \rightarrow & (\gamma,~g,~Z ~or~ \phi^0) ~+~ (c ~or~ u),
\end{eqnarray*}
have branching ratios less than $10^{-10}$ in the Standard Model.

\section{Searches}
Both CDF\cite{cdf} and D0\cite{d0} presented detailed results
on the search for top at the Tevatron.
The data presented included both the dilepton and the lepton plus jets mode
for the decay of the $t\bar{t}$ pair.
However, neither collaboration presented data
on the six jet mode. Theoretical calculations suggest that with an efficient
$b$-quark tag, that this mode will be accessible at the Tevatron.
A detailed summary of the experimental results present can
be found in the review of hadron collider physics by Shochet\cite{ms}.

CDF observes a $2.8~\sigma$ (0.26\%) effect which is not sufficient
to firmly establish the existence of top but which, if interpreted
as top, yields $m_t ~=~ 174 \pm 10 ^{+13}_{-12}~GeV/c^2$ and
$\sigma_{t\bar{t}} ~=~ 13.9 ^{+6.9}_{-4.8} ~pb$.

D0 does not observe a significant excess of events due to $t\bar{t}$
production.
The probability for the background to fluctuate to give greater than
or equal to the observed number of events is
7.2\% ($1.5~\sigma$). If $m_t ~=~ 180~GeV$
then $\sigma_{t\bar{t}} ~=~ 6.5 \pm 4.9 ~pb$.

Orr\cite{lo} presented the results of a study on the effects of soft
gluon radiation in the determination of the top quark momentum.
The results of this study will be important for precision measurements
of the top quark mass at hadron colliders.

\section{Surprises}

In the Standard Model the couplings of the top quark to each
of the gauge bosons, $g, \gamma, ~Z$ and $W^{\pm},$
are determined, including radiative corrections.
Therefore potential new physics could show up as deviations of these
vertices from Standard Model expectations.
Kao\cite{ck} and Rizzo\cite{tr} discussed corrections to
the QCD coupling, $gt\bar{t}$.
Kao's paper concentrated on the one-loop
weak corrections in both the Standard Model
and the Minimal Supersymmetric Standard Model.
Whereas Rizzo considered the effects of an anomalous chromomagnetic moment
to this coupling.

Schmidt\cite{bs} summarized top quark physics at $e^+e^-$
colliders and in particular discussed the signatures
of deviations to the $\gamma t\bar{t}, Z t \bar{t}$ and $W t \bar{b}$
vertices at such machines.

\begin{figure}[t]
\vspace{8cm}
\includegraphics{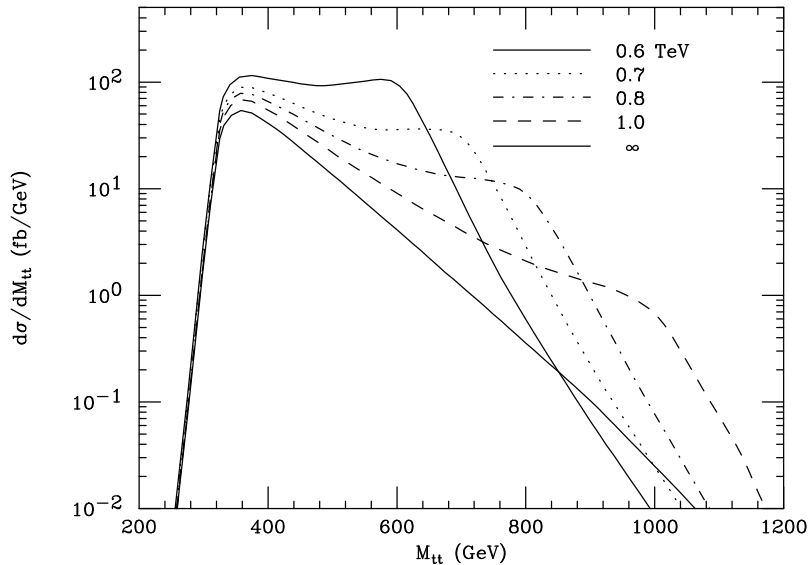}
\vspace{-1cm}
\caption[]{
The differential distribution, $\frac{d\sigma}{dM_{t\bar{t}}}$ verses
$M_{t\bar{t}}$ for top-antitop production with the curves labelled by the
mass of the octet top-color vector boson.
}
\label{mtt}
\end{figure}

Of course the top quark could present more dramatic surprises such as
charged Higgs decays, $t \rightarrow H^+ ~b$,
large flavor changing neutral current decays,\\
$t \rightarrow (g,Z ~or~ \gamma) ~+~ (c ~or~ d)$
or enhanced production through a new resonance.
This latest possibility could occur either through the quark-antiquark
production
mode\cite{hp}, as expected in Top-color models of electroweak symmetry breaking
or via the gluon-gluon fusion\cite{el} as suggested by some Technicolor models.
In both cases the $t\bar{t}$ pair is produced by the decay of a heavy new
particle, Top-color Boson or Techni-eta, which distorts the $P_T$,
$\cos \theta^*$ and $M_{t\bar{t}}$ distributions from the
Standard Model expectation. Fig. 3 is the change in the shape of the
$M_{t\bar{t}}$ distribution in the Top-color Model discussed by Hill
and Parke\cite{hp}. We should think of top quark production as a new Drell-Yan
process probing extremely high mass scales, greater than 500 GeV.

\section{Conclusion}

The top quark is an exciting new window on very high mass scale physics.
While exploring the vista from this window we should be on the lookout for
any deviation from the Standard Model which will provide us with information
about that elusive beast, the mechanism of electro-weak symmetry breaking.
Because the mass of the top quark is very heavy,
this quark is the particle most strongly
coupled to the electro-weak symmetry breaking sector.
Therefore the deviations could be seen at zeroth
order or may require more subtle measurements.

What is needed is hundreds of top-antitop pairs as soon as possible.
Then, watch out for surprises at DPF'96!

The author thanks E.~Laenen for providing Figure 2.
Fermilab is operated by the Universities Research Association under
contract with the United States Department of Energy.


\end{document}